# Building Executable Secure Design Models for Smart Contracts with Formal Methods

Weifeng Xu [1], Glenn A. Fink [2]

[1] College of Public Affairs, University of Baltimore
1420 N. Charles Street, Baltimore, MD 21201, USA
`wxu@ubalt.edu`

[2] Cyber Security Group, Pacific Northwest National Laboratory
902 Battelle Blvd, Richland, WA, USA
`Glenn.Fink@pnnl.gov`

**Abstract**. Smart contracts are appealing because they are self-executing business agreements between parties with the predefined and immutable obligations and rights. However, as with all software, smart contracts may contain vulnerabilities because of design flaws, which may be exploited by one of the parties to defraud the others. In this paper, we demonstrate a systematic approach to building secure design models for smart contracts using formal methods. To build the secure models, we first model the behaviors of participating parties as state machines, and then, we model the predefined obligations and rights of contracts, which specify the interactions among state machines for achieving the business goal. After that, we illustrate executable secure model design patterns in TLA+ (Temporal Logic of Actions) to against well-known smart contract vulnerabilities in terms of state machines and obligations and rights at the design level. These vulnerabilities are found in Ethereum contracts, including Call to the unknown, Gasless send, Reentrancy, Lost in the transfer, and Unpredictable state. The resultant TLA+ specifications are called secure models. We illustrate our approach to detect the vulnerabilities using a real-estate contract example at the design level.

## I. Introduction

Smart contracts are self-executing contracts with the terms of the agreement between parties being directly written into lines of code. Smart contracts have interesting properties: (1) they are unbreakable agreements with predefined rules, (2) they exist across a distributed, decentralized blockchain network, and (3) they permit trusted transactions and agreements to be carried out among disparate, anonymous parties without the need for a central authority, legal system, or external enforcement mechanism [1]. Regardless of their appeal, smart contracts are software, and therefore, they may contain potential defects common to software, such as those arising from the complexity of the modeled system. These defects are either design flaws or implementation bugs. Smart contract design flaws are often inherited from business contracts or introduced during smart contract design. Smart contract software defects are introduced during implementation. After defective smart contracts been deployed to a production blockchain, they become very expensive to fix because they are unchangeable and decentralized. There is a well-known recent incident that is related to the both technical and business issues of smart contracts [2]. On June 17, 2016, a hacker discovered a smart contract error in the Ethereum Distributed Autonomous Organization (DAO), eventually forcing the entire currency to do a "hard fork" to erase the fraudulent transactions from the ledger, costing tens of millions of dollars in losses. The DAO hack was the first well-known incident caused by defects in smart contracts and exploited by malicious users. It is legally unclear how other blockchain communities will agree to resolve similar incidents in the future.

The goal of this research is to build secure models for smart contracts to detect security vulnerabilities at the design level. Specifically, we (1) employ state machines to model the behaviors of business contracts, (2) analyze well-known smart contract vulnerabilities in terms of state machines, and (3) propose and build vulnerability discovering mechanisms while formalizing state machines using formal methods. The contributions of the paper include (1) separated design concerns: using state machines to model the obligations and rights of business contracts in two levels: the capabilities of participating parties and the constraints of these capabilities defined by contracts, (2) building smart contract vulnerability resistance capabilities. The capabilities are built based on the well-known design by contract principle [3] [4] and the well-known low level attacked captured in state machines [5], (3) facilitating vulnerability detecting automation: the proposed design patterns and approach facilitate the vulnerability detecting automation by formalizing behavioral models in TLA+ [6].

The rest of the paper is outlined as follows: Section II describes the overall approach. Section III presents the secure model design for smart contracts. Section IV reviews the related work. Finally, Section V concludes this paper.

## II. Background and Approach Overview

In this section, we briefly discuss the life cycle of smart contracts, review several well-known smart contract vulnerabilities, and then show the overall approach to building secure design models for smart contracts.

Smart contracts are software, and therefore, we should manage the software development lifecycle to minimize the chance of introducing defects. The key activities in the life cycle include (1) specifying business contracts, (2) designing smart contracts, (3) implementing smart contracts, (4) auditing smart contracts, and (5) deploying and executing smart contracts. Mistakes can be made in each activity caused by problems like the ambiguity of business contracts, software specification and design flaws, implementation errors in smart contract code, incomplete auditing, and environmental errors from the smart contract execution platform. We focus on the second activity of the smart contract development life cycle and aim at discovering the design flaws of smart contracts. As examples to guide the development of our approach, we are interested in discovering the following system behavior-related vulnerabilities mentioned in [7] [8]:

- Call to the unknown (CTU). Some of the primitives used in Solidity [9] to invoke functions and to transfer ether may have the side effect of invoking an undesirable behavior of the recipient.

- Gasless send (GS). When using the function send to transfer ether to a contract, it is possible to incur in an out-of-gas exception.

- Reentrancy (RE). A behavior may be unintentionally executed more than once. This class of flaw was the cause of the DAO's problems.

- Lost in transfer (LIT). If some ether is sent to an orphan address, it is lost forever.

The main idea of securing smart contract design is to build security models around business contracts against the vulnerabilities of smart contracts including those mentioned. Specifically, we aim at designing and specifying the countermeasure mechanisms, i.e., secure design patterns, in TLA+ based on behaviors of the contracts. The newly created models are called Temporal Logic of Actions (TLA) security models, and they are the formal specification of state machines with the integrated vulnerability detection ability. TLA is a shorthand for referring to the TLA+ specification language [6]. The rich language has well-defined semantics and was designed for expressiveness and ease of formal reasoning. The center of Figure 1 represents secure models of smart contracts in TLA+. The creation of the secure models are the results of a sequence of modeling activities. The dashed links indicate the modeling activities are not enforced. However, each activity results in a partial TLA model. The activities are:

- Eliciting business needs. This step captures business requirements and agreements in contractual use-case templates. The basic elements required for the agreement to be a legally enforceable contract include the offer, parties, consideration, terms and conditions, and acceptance. This step also highlights the relations among these elements.

- Modeling participating party behaviors. This step describes the behaviors of each party involved in the business contracts in the corresponding state machines.

- Modeling contractual rights and obligations. This step models the interactions among participating parties. These interactions are enforced by the mutual agreements.

- Secure smart contract design. This step secures smart contract design by implementing secure design patterns to against well-known smart contract vulnerabilities. The design process focuses on what the smart contract should do rather than on how it should do it. For example, a specification may say that a vending machine must dispense a soda if took a one-dollar bill (the *what*), without specifying the mechanism used to take dollar bills and vend soda cans (the *how*).

Note that the TLA security models can be verified using a Temporal Logic model Checker (TLC). TLC is a model checker for debugging a TLA model by checking invariance properties of a finite-state model of the specification.

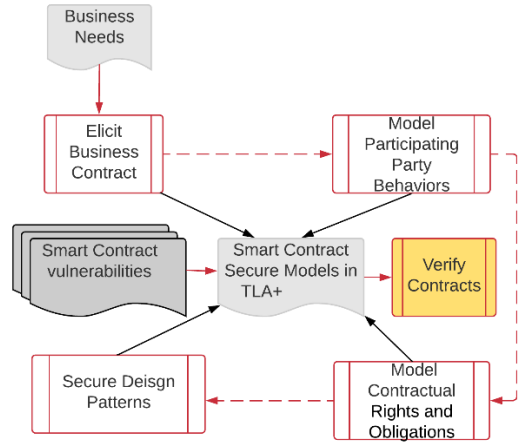

Figure 1. The systematic approach to build secure design models for smart contracts

## III. Modeling Business Contracts

In this section, we discuss the elicitation of the business contracts and model the behaviors of participating parties, including the obligations and rights of the participating parties in state machines.

### A. Eliciting Business Contracts

To demonstrate the approach, we use the example where an owner wants to sell his/her house and a potential buyer wants to purchase it. A property manager, such as a city where the house is located, will coordinate the transaction. An offer is the beginning of a contract. One party must propose an arrangement to the other, including definite terms. The *parties* of a contract are the entities involved in the agreement. For simplicity, we conflate the buyer/seller parties with their respective agents. The *consideration* of a contract specifies what each party stands to gain from the business arrangement. For example, if a house owner wants to sell his house to a buyer, the offer is the house, the seller's consideration is the payment for the sold house, and the buyer's consideration is ownership of the house. The *terms and conditions* of a business contract specify the rights and obligations of each party. These can vary widely depending on the nature of the business arrangement. Common examples can include the amount of payment, when payment is due, the specific nature of the work involved and how long the agreement will remain in effect. Terms and conditions may cover complex situations such as failure of one party to accomplish something required in a timely manner or dissatisfaction with the outcome of the contract resolution. *Acceptance* of the contract is the expression of assent to its considerations.

Business contract elicitation is the practice of collecting the contractual requirements from participating parties and other stakeholders. The contractual requirements are captured in contractual use-case specifications, which are an extension of use case specifications [10]. Specifically, a contractual use case's specification is a textual model that organizes and describes key components of business contract requirements. Table 1 shows a contractual use case specification based on the real-estate property-selling contract. It contains the key elements of business contracts, including a short description of the contract, an offer, the participating parties in the contract, the terms and conditions specified in the contract, the consideration of the contract, and the acceptance of the contract.

Table 1. A business contractual use case

| Name | Real Estate Property Selling |
|---|---|
| Description | A buyer and a seller want to exchange the ownership of a building after the buyer bought the building from the seller. |
| Offer | A house |
| Parties | A buyer, a seller, and a property manager |
| Legality | 1. The seller owns the house<br>2. The buyer has the ability to pay |
| Consideration | 1. The seller: payment for the sold house<br>2. The buyer: title deed to the house |
| Acceptance | 1. The seller: receipt of payment of the sold house<br>2. The buyer: receipt of ownership of the house<br>3. Property manager: the record of changed ownership |
| Rejection | 1. The seller: keep the house<br>2. The buyer: keep the money |
| Rights | See Table 2 |
| Obligations | See Table 2 |

Note that:

- The property manager acts as a contract manager. It specifies the business process and coordinates the behaviors of the both buyer and seller. The purpose of the property manager is to get rid of the middle man in the business process, which is one of the advantages of smart contracts. Each contract needs to define the contract manager.

- The use case template also includes the item "rejection" of acceptance, which describes a scenario that occurs when a participating party does not agree to the terms of the contract.

- Rights and obligations must be specified in terms and conditions of a business contract. Contractual rights are the set of rights guaranteed whenever people enter into a valid contract. For example, both parties have the rights to join the business negotiation process. After joining, both parties have obligations to inform the property manager regarding their status.

- The rights and obligations in the modeling process are considered as the interface among all parties in a contract. The interactions occurred in the interface.

## B. Modeling Contractual Rights and Obligations

Rights and obligations are the key elements of business contracts. We use state machines [11] to model the essential element. A state machine is an implementation-independent model of the dynamic behavior of the system. It is an abstract program representing a business contract that can be in exactly one of a finite number of states at any given time, involving all parties. A state machine can change from one state to another in response to some external inputs. The event-response features of the state machine represent the nature of obligations and rights between and among parties. Figure 2 shows the state machine of the rights and obligations defined in business contracts. The state machine models the behavior of the rights and obligations at two levels: The capabilities of each participating party expressed in state machines and the constraints of these capabilities.

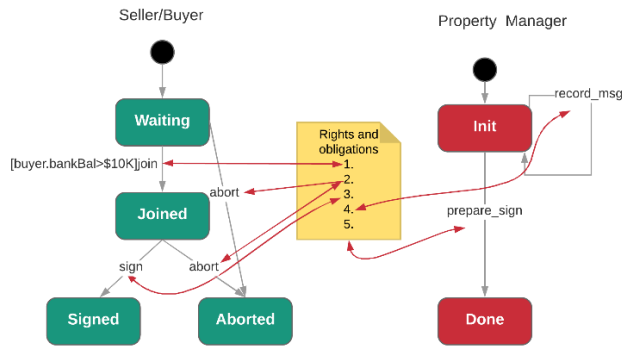

Figure 2. Rights and obligations between participating parties

**Modeling the capabilities of each party**. These parties include the seller, the buyer, and the property manager. The capabilities of a party describe what the party is capable of doing regardless of external constraints and stimulates. For example, the state machines of the seller and buyer show both parties are in the initial *"Waiting"* state, i.e., waiting for a contract. Both parties can abort the business process anytime. Once they have an intention to create a legal relationship, they can enter the *"Joined"* state indicating they have joined the business negotiation. Both seller and buyer have two options after joining the business negotiation either decide to sign the property selling contract or abort the business negotiation process. The outcome of each party is either *"Signed"* or *"Aborted"*. For the purpose of demonstration, the buyer and seller have separate instances of the same state machine. When the property manager changes his state from *"Init"* to *"Done"* he finishes preparing contractual documents and asking both seller and buyer to sign the contract. Note that the capabilities may have optional internal constraints. For example, to join the real-estate business, the buyer may need to show his/her bank account balance is greater than $10K. The internal constraint for that capability may be expressed as [buyer.bankBal>$10k]join. The property manager is often called the third-party beneficiary in business contracts.

**Modeling the constraints of these capabilities**. Although state machines capture each individual participating party's capabilities in a business setting, they do not describe the constraints of capabilities to achieve a business goal of a contract. Such constraints need to be specified as the rights and the obligations of each party. Rights are the fundamental normative rules about what parties are allowed to do. A contractual obligation means that a person must comply with the directives stated or given due to the agreement, promise, or verbal/written contract that is in place between the individuals involved. In the modeling process, the performance of obligations is formed from a party's responses (i.e., capabilities) to external events under the predefined terms and conditions (i.e., external constraints). *The external events* are the events generated by the status-changing information shared between the parties. The changes of such statues often have impacts on the offer or the process of making and accepting the offer. For example, if the offer of the contract is a house, then external events can be a price-change event, a time-change event, and message-receive event. An *external event flow* refers to the associate and the direction between the parties who generate and receive the events. The external event flows between the parties, and they are represented by red lines in Figure 2.

Figure 3 shows an example model of how an obligation is performed and a right is requested by the property manager. The model views the obligations and their corresponding rights as a three-step process:

- Obligation request. An obligation is requested by a smart contract. The dashed lines indicate the requested obligation process.

- Obligation performance. When an obligation is performed, the capability of a corresponding party is invoked.

- Right request. After an obligation is performed, the party can request his rights based on the contract. The solid links represent the rights requesting process.

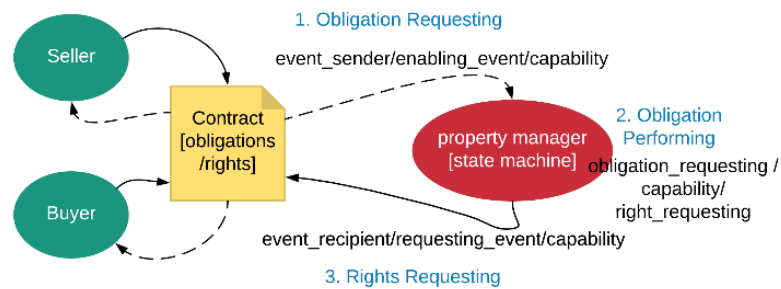

Figure 3. The model of performing an obligation

When modeling and integrating contractual rights and obligations of a party into state machines, we need to answer the following questions that are related to the obligations and rights: (1) who requested the services, (2) who provides the services or performs obligations, and (3) which obligation should be performed. We define the obligations of a party as the capabilities of the participating party invoked by predefined external events, where the capabilities are the behaviors captured by state machines. The external events can be generated by either receiving a message, a timer, or the changed states of globally shared variables by all participating parties. In other words, the obligations enable response to external *enabling events*. The rights of a party are obligation request events generated by the party. The requesting events ask other parties to provide pre-defined services in contracts in exchange for the party's services. We call these *requesting events*. The complete obligation-request and right-request messages are expressed as *event_sender/enabling_event/capability* and *event_recipient/requesting_event/capability,* respectively. The obligation performance is expressed as *obligation_request/ capability/rights_request*. It indicates the capability is invoked if an obligation request is received and the rights request will be fired. These expressions answer the three questions mentioned earlier.

Table 2 shows the rights and obligations of the seller and buyer. For example, once the property manager receives "*joined_msg*" messages (i.e., an enabling event) from both the buyer and the seller, the property manager is obligated to prepare the signing process (i.e., "*prepare_sign*"), and then sends a sign message "*sign_msg*" (a requesting event) to both the buyer and seller to sign the contract. It worth noting that we model shared attributes among parties, and the external events are generated based on the predefined rules consisting of the shared attributes.

## IV. Secure Smart Contract Design

We will discuss the design by contract principle and how the design patterns use the principle to detect vulnerabilities at a design level, including the misbehaviors of individual participant party, and the obligations and rights vulnerabilities among parties.

Table 2. The rights and obligations of each party

| ID | Parties | Obligations | | | | | Rights | |
|---|---|---|---|---|---|---|---|---|
| | | **Event Sender** | **Enabling Events** | **Capability** | | | **Requesting Events** | **Event Recipient** |
| | | | | **Current State** | **Activities** | **Next State** | | |
| 1 | Buyer /Seller | n/a | n/a | Waiting | Join the real estate process (*join*) | Joined | *joined_ msg* | property manager |
| 2 | Buyer /Seller | n/a | n/a | Waiting or Joined | Abort the transferring process any time before signing the transferring documentation (*abort*) | Aborted | | |
| 3 | Buyer /Seller | property manager | *sign_msg* | Joined | *sign* the selling/buying agreement | Signed | | |
| 4 | Property manager | buyer/seller | a message *joined_msg* | Init | Record the received messages (*record_msg*) | Init | | |
| 5 | Property manager | buyer and seller | *joined_msg* | Init | Prepare to sign (*prepare_sign*) | Done | *sign_msg* | buyer and seller |

## A. Design by Contract and TLA+

The state machine captures the capabilities of parties and contractual relations among parties at the design level, however, it doesn't enforce prevention of potential vulnerabilities. We secure the state machines by adding logic for secure design patterns that address well-known attacks on state machines, including extra states, sneak paths, trap doors, and the hidden activities [5]. Secure smart contract design is based on the well-known design-by-contract principle [3] [4]. The principle suggests that a software component must specify formal, precise, and verifiable interfaces by defining the component's preconditions, postconditions, and invariants. These specifications are also referred to as "contracts", in accordance with a conceptual metaphor with the terms and conditions and obligations of business contracts, but they must not be confused with the smart contract we are implementing.

We use TLA+ to secure smart contract design because TLA+ uses mathematics in a simple, native way to specify state machines [12]; it facilitates the implementation of secure design patterns and automates vulnerability detection. TLA+ has demonstrated the value of formal methods for Real-world Systems. Since 2011, engineers at Amazon have been using TLA+ to help solve difficult design problems in critical systems [13] [14].

## B. TLA+ Secure Design Patterns for Each Party's Behaviors

**Extra state checking design pattern.** The existence of extra states indicates the ability of parties to enter a situation unanticipated by the design. To prevent extra states, TLA+ allows us to define predicate logic formulas to check the type-correctness invariant. The formula creates a "positive" security model (also known as "whitelist") that defines what is allowed, and rejects everything else [15]. For example, lines 5, 6 and 7 of the TLA+ snippet below defines a formula named "checkExtraStateVul", which requires that the resultant state in the buyer/seller state machines must be one in the set {"Waiting", "Joined", "Signed", "Aborted"}.  Similarly, any property manager state must in the set {"Init", "Done"}. The symbol /\ in TLA+ is the propositional logic and operator and the symbol \in is a set operator ∈.  In addition, the symbol [] defines a function to be a set of ordered pairs. The expression [S -> T] is the set of all functions f whose domain is S such that f [x] is in the set T for all x in S. In our example, SB is the domain. The expression [SB -> { "Waiting", "Joined", "Signed", "Aborted" }] at line 6 is the mathematically way of listing all possible states of the buyer and the seller using functions, e.g., one of the ordered pairs can be (buyer |-> Waiting), where the symbol |-> separates the function domain and the range.

```
1.    CONSTANT SB      \* The participating parties, i.e., seller and buyer
2.    VARIABLE
3.        sbState,     \* The state of the seller and the buyer
4.        pmState      \* The state of the property manager
5.    checkExtraStateVul ==
6.      /\sbState \in [SB -> { "Waiting", "Joined", "Signed", "Aborted" }]
7.      /\ pmState \in {"Init", "Done"}
```

**Sneak path & trap door checking design pattern.** Once we eliminate unknown states we must secure the transitions between the legal states. A sneak path is an activity or an event that causes a transition out of a legal state under conditions that are not allowed. A sneak path may be triggered by incorrect conditions of the activity or when the current state of an instance is out of synchronization with the rest of the model. For instance, a sneak path might cause the property manager to enter its "Done" state prematurely by fooling it into erroneously believing that both the seller and buyer have signed. While all the instances are in legal states, the sneak path produces a system that is in an illegal meta-state.

A trap door is when a legal state of an implementation accepts an event that is not defined in the specification, which causes the instance to enter that state under unspecified conditions. To prevent sneak paths and trap doors, we whitelist the activities by adding additional constrains. Specifically, the secure design pattern specifies four constraints for each activity: (a) the current state, (b) the conditions of the activity, (c) the resultant state, and (d) the states of other participating parties, where the constraints (a), (b), and (c) addresses sneak paths, and constraint (d) addresses the trap doors. The following TLA+ design pattern specifies the constrained activities, and therefore, prevents the sneak paths.

```
1.  activity ==
2.       /\ current_state                    (a)
3.       /\ conditions                       (b)
4.       /\ next_state                       (c)
5.       /\ states_of_other_parties          (d)
```

The following TLA+ snippet defines a *sign* formula. Lines 2 and 3 check sneak paths, which implement the constraints (a) and (b). Line 4 checks any violations of trap doors, which implements the constraints (c) and (d). Line 4 means that the current state of the participating party must be "*Signed*" and states of other participating parties must remain unchanged. For the purpose of designing secure models, we do not implement the "*sign*" activity, and the formula is only to specify what the valid "*sign*" behavior is. If we want to model whether a buyer can sign the contract, we can introduce a `balance` variable and simplify assert the buyer's bank account has enough balance use the formula `canSign == balance >= propertyPrice`.

```
1.  sign(sb) ==
2.       /\ sbState[sb] = "joined"
3.       /\ canSign
4.       /\ sbState' = [sbState EXCEPT ![sb] = "signed"]
```

**Hidden activity checking design pattern.** Hidden activities are illegal activities that should not be allowed by the parties. To prevent the existence of hidden activities, we can simply use TLA+ to list only specific activities are allowed in the contracts. The following TLA+ snippet defines only three activities *"join", "sign",* and *"abort"* allowed for the seller and buyer. The symbol \/ in TLA+ is the propositional logic or operator

```
1.   Next == \E sb \in SB : join(sb) \/ sign(sb) \/ abort(sb)
```

## C. TLA+ Secure Design Patterns for Obligations and Rights

The secure design pattern (shown below) for obligations and rights requires the following formulas: the attributes shared by the parties, shared_attributes, the allowed external_events, an obligation, a right, a term_condition, a post_status, and an instance of each state machine SM_inst.

```
1.  shared_attributes
2.
3.  external_events==[?] \* enabling events and requesting events
4.
5.  Term_condition ==
6.      /\ obligation
7.      /\ right
8.      /\ post_status
9.
10. obligation ==
11.     /\ event_sender
12.     /\ enabling_events
13.     /\ current_state
14.     /\ next_state
15.
16. right ==
17.     /\ event_recipient
18.     /\ requesting_event
19.
20. post_status==?
21.
22. SM_inst == INSTANCE partyStateMachines
```

The secure design pattern has two purposes: (a) mapping key elements of business models, particularly the rights and obligations, and (b) detecting the aforementioned vulnerabilities, including GS, RE, LIT, and CTU:

- Detecting GS vulnerability. GS vulnerability is essentially a sneak path. The vulnerability is related to each party's behaviors rather than obligations and rights. The formula SM_inst indicates that the behaviors of each individual party described in the previous section must be verified when checking the rights and obligations of parties. SM_inst has defined the formula conditions to enforce the conditions before carrying out any activities, and therefore, it prevents any sneak paths activated by incorrect conditions.

- Detecting RE vulnerability. RE vulnerability is often caused by a sneak path or a trap door between parties. To detect RE vulnerability at the design level, the secure design pattern adds additional constraints, including current_state, next_state, event_sender and enabling_events, in the obligations formula. The obligation formula indicates that (1) the first entry needs to reach the correct next_state, (2) the second entry needs to meet the current_state, and (3) an obligation is valid only if the required enabling_events are generated by the predefined event_sender party and the events are predefined in the external events set external_events. It prevents sneak paths from unspecified scenarios and trap doors between participating parties because of undefined external events or incorrect event senders. Note that the formula external_events includes both types of external events: enabling events and requesting events. The formula asserts that any events that are not in the set are invalid. These external events are generated based on the shared attributes or by other parties directly. In addition, if the RE vulnerability involves two parties, we use the secure design for contract considerations patterns (discussed in next section) to enforce the integrity of the contract.

- Detecting LIT vulnerability. SM_inst defines CONSTANT SB, which specifies the parties in the contracts. It only allows transferring assets between parties. Furthermore, the vulnerability can be detected by verifying both considerations of the seller and buyer, e.g., the formula ContractConsistentCheck, which models the mutual states between two parties.

- Detecting CTU vulnerability. CTU triggers side effects while fulfilling obligations and claiming rights. Two mechanisms are proposed to reduce the vulnerability: preventing the incorrect invocations and post-statues checking. For example, to prevent claiming unknown rights, the secure design pattern has two formulas: (a) use a right formula to prevent generating incorrect events, and (b) use a post_status formula to detect the side effect of claiming incorrect rights. The rights of a participating party can be claimed only if a requesting event requesting_event is generated and the right event_recipient is defined. The post_status formula asserts what states must remain unchanged to prevent side effects.

The following TLA+ design snippet specifies the fifth right and obligation listed in Table 2. Each formula is described as follows:

We first define an event set (line 1). This specifies two types of messages that are allowed to pass between the seller and the buyer: a *"joined_msg"* message (line 2) sent by the seller or the buyer and a *"sign_msg"* message (line 3) sent by the property manager. Other events are not in the set may lead to GS vulnerability. Note that the special function in line 2 [type : {"joined_msg"}, sb : SB] represents a set of records with a domain named as type and a range named as sb. The \cup symbol indicates the union operation.

Lines 12-15 specify obligations. It indicates that the obligation is valid only if the property manager receives joined messages from the buyer and seller (line 13) the current state of the property manager, *"pmState"*, is *"Init"* (line 14), and the next state of the property manager is *"Done"* (line 15).

```
1.  events ==
2.      [type : {"joined_msg"}, sb : SB] \cup
3.      [type : {"sign_msg"}]
4.
5.  VARIABLES msgs_passing, pmJoinedMsgsReceived
6.
7.  pmSign ==
8.      /\ obligation
9.      /\ right
10.     /\ post_status
11.
12. obligation ==
13.     /\ pmJoinedMsgsReceived = SB
14.     /\ pmState = "Init"
15.     /\ pmState' = "Done"
16.
17. right == msgs_passing '
18.     = msgs_passing \cup {[type |-> "sign_msg"]}
19.
20. post_status == UNCHANGED <<sbState, pmJoinedMsgsReceived >>
```

Lines 17-18 specify rights of the property manager. It changes the states of the shared attributes "*message_passing*" by adding a "*sign_msg*" to the attributes. The added message will trigger the third obligation of both buyer and seller that is defined in Table 2. Line 20 specifies that the obligation and right should not change the states of both the buyer and the seller as well as the variable that the property manager used to hold the received messages from the buyer and the seller. This prevents exploitation of the CTU vulnerability.

### C. TLA+ Secure Design for Contract Considerations

Considerations are the benefits that each party receives, or expects to receive when entering into a contract. For a business contract to be considered valid and enforceable by the courts, three elements of consideration must be met: (a) there is a bargain for the terms of the exchange, (b) the bargain includes a mutual exchange between the parties, and (c) the exchange is something of value. Secure design patterns for contract considerations include a model of each party's considerations and a model of the mutual exchange.

Each party's consideration can be modeled in terms of value consideration changes shown below:

```
1.  considerCheck (party)==
2.      /\current_state
3.      /\valueConsider'=valueConsider+ consideration
4.      /\next_state
```

For example, if the seller's consideration is the payment received, the secure design pattern needs to verify the payment has been received in the state model. The third line in the activity "*sign*" verifies that if the seller sold his property, his/her bank account balance will be increased by the payment of the property.

```
1.  sign(seller) ==
2.      /\ sbState[seller] = "joined"
3.      /\ balance'= balance + payment
4.      /\ sbState' = [sbState EXCEPT ![seller] = "signed"]
```

Modeling the mutual exchanges between two parties. It takes both parties' considerations together as a whole. It often models the final states of both parties and verify a value flow between two parties after mutual exchange of a consideration. The following formula asserts that the seller and the buyer should not have arrived at conflicting decisions. It guarantees that either both parties sign the contract or both of them abort the business process.

```
1.  ContractConsistentCheck ==
2.  \A sb1, sb2 \in SB :
3.      ~ /\ sbState[sb1] = "aborted"
4.        /\ sbState[sb2] = "signed"
```

The following formula `ContractPaymentConsistentCheck` asserts that the payment transfers from the buyer to the seller. Similarly, we can verify the ownership transferring.

```
1.  ContractPaymentConsistentCheck ==
2.  \A sb1, sb2 \in SB :
3.      /\ seller_sign(sb1)
4.      /\ buyer_sign(sb2)
5.
6.  seller_sign1(seller) ==
7.      /\ balance[seller]'= balance[seller] + payment
8.
9.  buyer_sign(buyer) ==
10.     /\ balance[buyer]'= balance[buyer] - payment
```

## V. Related Work

Smart contract verification concepts are not new. One of the early works is done in 1997 by Szabo [16]. Szabo described the basic idea behind smart contracts as different kinds of contractual clauses (such as collateral, bonding, delineation of property rights, etc.). These contractual clauses can be embedded in the hardware and software we deal with, in such a way as to make a breach of contract expensive for the cheater. Szabo used protocols and user interfaces to formulate all steps of the contracting process. This work provides new primitives to formalize and secure digital relationships. Grosof et al. [17] [18] built a rule-based approach to the representation of business contracts that enables software agents to create, evaluate, negotiate, and execute contracts with substantial automation and modularity. It builds upon the situated courteous logic programs knowledge representation in RuleML. Similarly, Governator [19] presented an approach for the specification and implementation of e-contracts for Web monitoring. This is done in the setting of RuleML.

He argued that monitoring contract execution requires also a logical account of deontic (rule-based) concepts and of violations.

Smart contract verification for blockchains is relatively new, however, there is a large body of similar work on formal software verification. Bhargavan et al. [20] outlined a framework to analyze and verify both the runtime safety and the functional correctness of Ethereum contracts by translation to F*, a functional programming language aimed at program verification. Their approach is based on shallow embeddings and type-checking within an existing verification framework. It does not address specific smart contract vulnerabilities. Delmolino et al. [21] documented several typical classes of mistakes students made, suggest ways to fix/avoid them and advocate best practices for programming smart contracts. Their work mainly focused on discovering bugs at the code level. Bigi et al. [22] combine game theory and formal models to tackle the new challenges posed by the validation of such systems. They extends Markov Decision Process to model the behaviors of the participants.

The proposed approach is rooted in two concepts in software engineering: design by contract and TLA+ formal methods. The central idea of design by contract is a metaphor on how elements of a software system collaborate with each other on the basis of mutual obligations and benefits.

## VI. Conclusion

The design of smart contracts needs to be checked and verified to minimize the design flaws and detect security vulnerabilities. We have presented a systematic approach to build secure models for smart contracts in TLA+ to verify the smart contract design. We have applied the approach to a property sale sample contract. Specifically, we have demonstrated how TLA secure models are generated to address some well-known smart contract vulnerabilities, including GS, RE, LIT, and CTU. This approach models the elements of business contracts in state machines and propose secure design patterns in TLA to detect smart contract vulnerabilities at the design level.

For future work, we plan to (1) extend the case study by increasing more states and behaviors to approximate a real-world scenario, (2) implant vulnerabilities in the property sale contracts and evaluate the vulnerability detection rates of the secure models, and (3) develop secure smart contract design templates so that the templates can be generated automatically to detect smart contract vulnerabilities. In addition, the template can help developers to cover general business informal contracts. These informal contracts exist in business contracts and do not require a specified form or method of formation in order to be valid. However, they may be required in smart contracts to reduce malicious behaviors.

ACKNOWLEDGMENT

This work is supported in part by the Department of Energy and the National Science Foundation under Grant Numbers 1714261.